\documentclass[aps,twocolumn,prl]{revtex4}
\usepackage{graphicx}

\newcommand{\gtsim}{\protect\raisebox{-0.5ex}{$\:\stackrel{\textstyle >}
        {\sim}\:$}}

\begin{document}

\title{Short-ranged attractions in jammed liquids: 
How cooling can melt a glass}

\author{Phillip L. Geissler}
\affiliation{Department of Chemistry,
University of California, and Physical Biosciences and Materials
Sciences Divisions, Lawrence Berkeley National Laboratory, Berkeley, CA
94720}
 
\author{David R. Reichman}
\affiliation{Department of Chemistry, Columbia University,
3000 Broadway, New York, NY 10027}
 
\date{\today} 
 
\begin{abstract}
We demonstrate that an extended picture of kinetic constraints in
glass-forming liquids is sufficient to explain dynamic anomalies
observed in dense suspensions of strongly attracting colloidal
particles.  We augment a simple model of heterogeneous relaxation with
static attractions between facilitating excitations, in a way that
mimics the structural effect of short-ranged interparticle
attractions.  The resulting spatial correlations among facilitated and
unfacilitated regions give rise to new relaxation mechanisms that
account for non-monotonic dependence of relaxation times on attraction
strength as well as logarithmic decay of density correlations in time.
These unusual features are a simple consequence of spatial segregation
of kinetic constraints, suggesting an alternative physical perspective
on attractive colloids than that suggested by mode-coupling theory.
Based on the behavior of our model, we predict a crossover from
super-Arrhenius to Arrhenius temperature dependence as attractions
become dominant at fixed packing fraction.
\end{abstract}

\maketitle

\newpage
Homogeneous liquids with slowly varying intermolecular attractions are
typically well represented by appropriate hard sphere reference
systems\cite{wca}.  The deeply supercooled regime of simple liquids
appears to be no exception\cite{ediger}.  Computer simulations of hard
sphere fluids at high packing fraction, $\phi$, exhibit the canonical
features of glass-forming materials, including dynamic heterogeneity,
stretched exponential decay of dynamical correlators, and dramatic
changes in relaxation times following small changes in density
(corresponding in most experiments to small changes in temperature
$T$)\cite{heuer}.  Confocal microscopy of colloids under conditions
designed to maximally screen interparticle attractions confirms this
fact\cite{weitz}.  Most of the theoretical approaches developed to
rationalize sluggish dynamics take advantage of the fact that
intermolecular structure at such high densities (or low temperatures)
differs little from that of the standard liquid state near the triple
point.  It is exactly this spatially uniform structure that justifies
a hard sphere representation.

Glass-forming liquids that do not fit into this van der Waals picture
thus not only introduce the possibility of qualitatively new dynamical
features but also provide a significant test of the flexibility of
various theoretical perspectives.  The simplest such material is a
fluid of hard spheres that attract one another strongly over short
distances, realized experimentally as a suspension of attractive
colloidal particles.  In this case attractive forces can play an
important role in determining intermolecular structure.  Experiments
and computer simulations indicate that for small $u/T$ (where $u$ is
the strength of attractions), overall relaxation rates exceed those of
a hard sphere liquid at the same packing
fraction\cite{poon,chens,sciortino,cates,eckert,sciortino3,pham}.  For
fixed $u$, one can thus view melting of the glass as a consequence of
cooling.  For large $u/T$, however, tight clustering of particles
leads to rigid gel-like structures that rearrange extremely slowly.
This second, attraction-driven glassy state would seem to differ
qualitatively from the original jammed material ($u=0$) in both
structure and dynamics\cite{sciortino2}.

Calculations based on mode-coupling theory have motivated much of the
experimental study of slow dynamics in attractive colloids, predicting
the existence of multiple glassy states and relaxation times that vary
non-monotonically with $u/T$ at fixed
$\phi$\cite{goetze,pusey,dawson}.  The physical origins of these
predictions, however, are difficult to assess.  Attraction-driven
changes in dynamics have been ascribed to loosening of the dense
liquid environment that confines particle motion.  But the mean field
approximation underlying mode-coupling theory would seem unable to
capture the inhomogeneous particle clustering necessary for such a
mechanism.  This mean field nature is most problematic in generating a
spurious dynamical arrest at large $u/T$ or $\phi$.  Nonetheless, many
predictions of mode-coupling theory, including logarithmically slow
relaxation near re-entrance to the attractive glass, have been
verified by dynamic light
scattering\cite{poon,chens,sciortino,cates,eckert,sciortino3,pham}.

A more intuitive, though less detailed, theory of deeply supercooled
liquids has been constructed by considering {\em only} the
inhomogeneous fluctuations that lie outside the scope of mode-coupling
theory\cite{harrowell,harrowell2,jpdc,jpdc2}.  Jamming at high density
prevents all but a small fraction of particles from moving a
significant distance over any short time interval.  The loose regions
(or defects) which enable nearby relaxation are therefore sparse and
essentially uncorrelated in space.  Although the statistics and
spatial patterns of these defects may be trivial, their dynamics are
complicated by the fact that jammed regions cannot spontaneously
loosen unless nearby motion provides space for them to do so.  In
other words, defects can only be created in the vicinity of other
defects.  Such a constraint greatly limits the accessible pathways for
exploring phase space\cite{jpdc}.

The simplest model embodying this picture consists of non-interacting
excitations whose dynamics are coupled together by kinetic
constraints.  Each cell of a lattice, with the size of a liquid's
density correlation length, is thus designated as either excited,
$n_i=1$, or jammed, $n_i=0$.  Creating defects in a liquid requires
local reorganization of particles and thus costs entropy (or, in a
thermal system, free energy).  This cost is represented schematically
by an external field, $h>0$, limiting the average concentration of
mobile cells, $\langle n_i \rangle \equiv c \ll 1$.  The model
system's total energy is then ${\cal H} = h \sum_{i=1}^N n_i$.
(Throughout this paper we express all energies in units of $T$,
distances in units of the lattice spacing, and frequencies in units of
the rate at which an unbiased, unconstrained cell changes its
excitation state.)  Cell mobilities change with rates that preserve a
canonical distribution, but are subject to a constraint of
facilitation.  Namely, cell $i$ is only free to change its state when
one of its nearest neighbors is already excited.  In the
one-dimensional East model only cell $i-1$ can facilitate changes at
cell $i$\cite{jackle}.  In the Fredrickson-Andersen model either
neighbor ($i+1$ or $i-1$) can provide facilitation\cite{FA}.

Despite their simplicity one-dimensional facilitation models exhibit
remarkably nontrivial dynamics.  Garrahan and Chandler have shown that
kinetic constraints in these models confer special geometries on
trajectories below a crossover value of $c$\cite{jpdc}.  These
spatio-temporal patterns result in sluggish relaxation with
temperature dependences mirroring those of many glass-forming
materials\cite{ritort,berthier,jpdc2}.  There are even indications
that scaling of such geometric features near a dynamical critical
point at $T=0$ may be universal among facilitation models and real
liquids\cite{whitelam}.  Nevertheless, a detailed relationship between
between mobility-promoting defects and molecular degrees of freedom
remains elusive.

It is not obvious that the kinetic facilitation picture is
sufficiently flexible to account for the anomalous dynamical behavior
reported for colloidal suspensions\cite{cates2}.  Basic facilitation
models possess a single control parameter, $c$, which seems best to
correspond to the density (or temperature) of a real material.  The
monotonic growth of relaxation times with decreasing $c$ in these
models appears to prohibit a straightforward explanation of
re-entrance and multiple glassy states in terms of dynamic
heterogeneities alone.  In this Letter we describe a simple extension
of the kinetic facilitation picture inspired by structural changes
arising from short-ranged attractions.  We then examine the dynamics
of an appropriately modified one-dimensional East model and
demonstrate striking agreement with the phenomenology of attractive
colloids.

We assert that the primary effect of introducing short-ranged
attractions in a dense, disordered material is to produce a gradual
segregation of tightly clustered, immobile regions from loose regions
enriched with facilitating defects.  In the limit of very strong
attractions such demixing produces gel-like structures and eventually
macroscopic phase separation, as seems appropriate for strongly
attractive colloids.  In the context of one-dimensional facilitation
models, gradual demixing is naturally generated by adding attractions
between adjacent excitations:
\begin{equation}
{\cal H} = -\epsilon \sum_{i=1}^N n_i n_{i+1} + h(\epsilon,c)
\sum_{i=1}^N n_i.
\label{equ:hamiltonian}
\end{equation}
Although clusters and voids would certainly segregate as desired with
increasing attraction strength $\epsilon$, the average concentration
of mobile cells would increase significantly if the external field $h$
did not change.  We imagine that increasing attraction strength in a
liquid with fixed density, however, does not significantly change the
concentration of facilitating defects (in a sense, the total free
volume).  If we view $c$ as uniquely determined by $\phi$, then the
field strength $h$ must be adjusted to maintain a desired value of $c$
for any $\epsilon$.  With a routine change of variables, $s_i = 2n_i -
1$, Eq.~\ref{equ:hamiltonian} becomes (within an additive constant)
the energy of a one-dimensional Ising model with exchange interaction
$J=\epsilon/4$, magnetic field $H=-(h-\epsilon)/2$, and magnetization
per spin $\langle s_i \rangle = -1 + 2c$.  The required adjustment of
external field given $c$ and $\epsilon$ is thus a standard
result\cite{imsm}:
\begin{equation}
h(\epsilon,c) = \epsilon + 2\sinh^{-1}\left({1-2c \over 2
\sqrt{c(1-c)} }\,\, e^{-\epsilon/2}\right).
\label{equ:field}
\end{equation}
This constraint on the density of excitations, motivated by the notion
of conserved free volume at a specific packing fraction, distinguishes
our approach from simply adding attractions to a conventional model of
kinetic facilitation.  It is an essential feature, without which
dynamical anomalies such as re-entrance would not be obtained.

\begin{figure}
\centerline{\includegraphics[width=8cm]{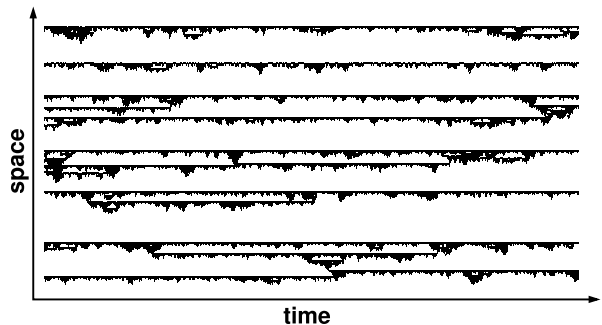}}
\caption{A representative trajectory of the one-dimensional attractive
East model with $c=0.10$ and $\epsilon=4.0$.  Vertical cross-sections
are snapshots of a portion of the system (several hundred lattice
cells), in which excited lattice cells are colored black.  Unexcited
cells are not shown.  Time runs along the horizontal axis. The total
duration of this trajectory segment is $1000$ units of the relaxation
time of a single facilitated cell.  See Ref.~\cite{jpdc} for a
detailed discussion of this manner of depicting trajectories and the
space-time geometry it reveals.}
\end{figure}

We emphasize that the attraction between defects in our model does not
represent a literal interaction energy between mobile regions in a
liquid.  In a liquid it is of course particles that attract one
another, the effect of which is to drive mobile regions together.
Attractions in Eq.~\ref{equ:hamiltonian} could thus be viewed as a
potential of mean force between mobile regions, or simply as a measure
of clustering tendency.  For a fixed value of $c$, it is in fact
energetically immaterial whether attractions act between adjacent
excited lattice cells, $n_i$, or between adjacent jammed cells,
$1-n_i$.

The structural effect of nonzero $\epsilon$ for a particular value of
$c$ is clear: Excitations cluster together, and the average spacing
between consecutive clusters, $d$, grows:
\begin{equation}
d \approx \cases{
d_0 + (e^{\epsilon} -1), & $\epsilon \ll h_0$\cr
c^{-1/2}e^{\epsilon/2}, & $\epsilon \gg h_0$
.\cr}
\end{equation}
Here, $d_0\approx c^{-1}$ and $h_0\approx \ln(1/c)$ are the values of
$d$ and $h$ at $\epsilon=0$.  Just as short-ranged attractions enhance
transient structural heterogeneity in a liquid, attractions in the
East model generate nontrivial (i.e., non-ideal gas) spatial patterns
of defects.  The dynamical consequences are more subtle.  In a
Metropolis Monte Carlo trajectory the rate of exciting a cell at the
boundary of an excitation domain (or alongside a single excitation) is
$e^{-(h-\epsilon)}$. Larger values of $J$ in the isomorphic
Ising model require weaker symmetry-breaking magnetic fields to
maintain fixed magnetization, so that $h(\epsilon,c)-\epsilon$
decreases monotonically with $\epsilon$:
\begin{equation}
h-\epsilon \approx \cases{
h_0 -\epsilon + 2c(e^{\epsilon} -1), & $\epsilon \ll  h_0$\cr
c^{-1/2}e^{-\epsilon/2}, & $\epsilon \gg h_0$
.\cr}
\end{equation}
Attractions therefore expedite fluctuations in the vicinity of
excitations.  This enhanced defect creation rate must be balanced,
however, in order to preserve the proper equilibrium distribution.
Here, balance is provided by the decline in number of excited clusters
per lattice cell.

In the FA model defects may diffuse freely through the system through
short sequences of local moves, e.g., $10\rightarrow 11 \rightarrow
01$.  The directionality of constraints in the East model gives rise
to more complex, self-similar relaxation pathways when the density of
excitations is low.  In this case all excitation domains are pinned on
one side, as is clear from the trajectory plotted in Fig.~1, and can
fully relax only when contacted by another domain on that side.  For
$\epsilon=0$, the minimum energy path for connecting two defects
separated by a distance $\ell=2^m$ establishes a set of isolated
defects between them.  These stepping stones are situated
hierarchically, at $\ell/2, 3\ell/4, 7\ell/8, \ldots, \ell-1$, giving
rise to an energetic barrier $\Delta(\ell) = mh_0$\cite{evans}.  The
overall relaxation rate is then $\Gamma_0 \approx e^{-\Delta(d_0)} =
d_0^{-h_0/\ln{2}}$.

\begin{figure}
\centerline{\includegraphics[width=8cm]{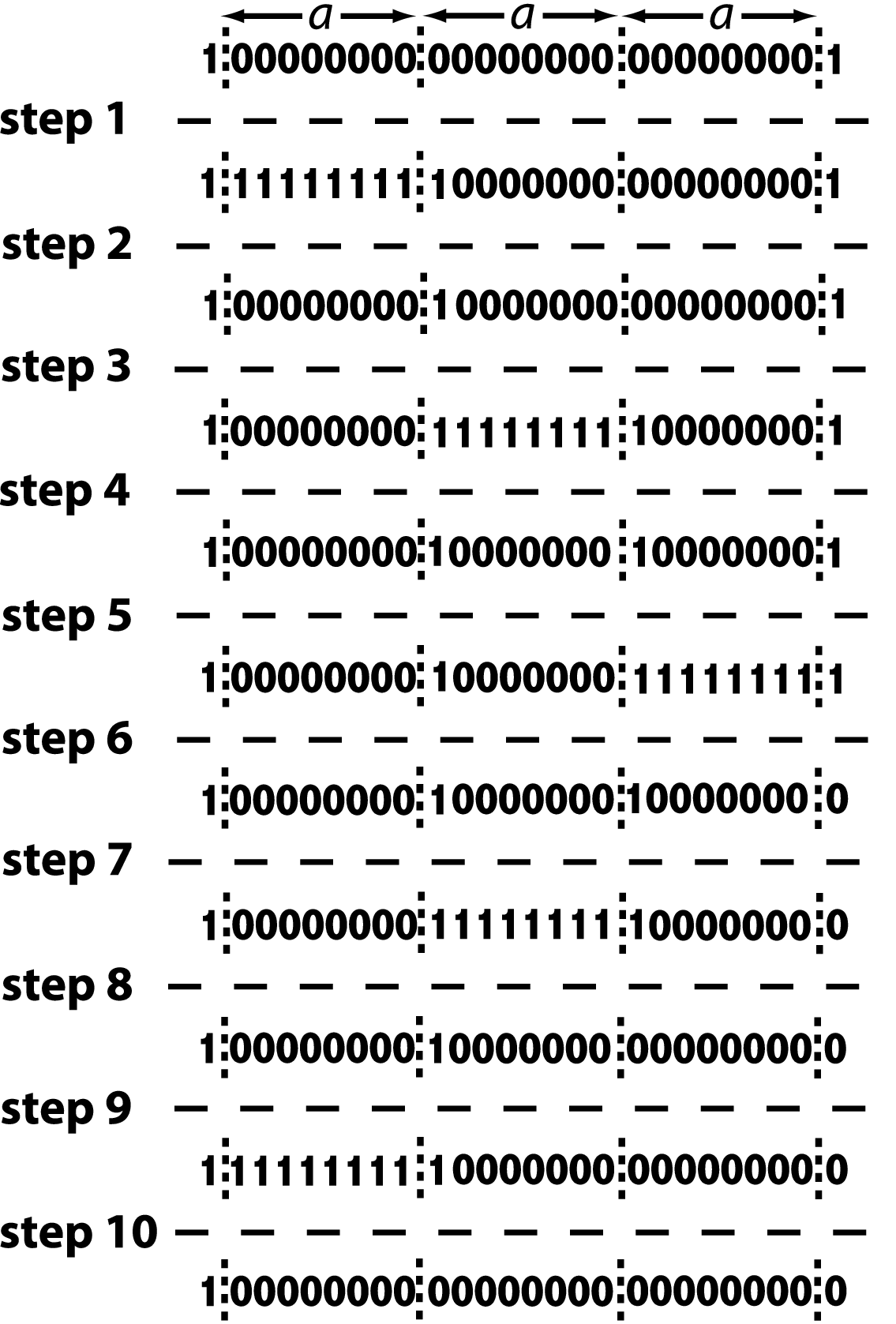}}
\caption{Renormalization of the hierarchical relaxation mechanism
characteristic of the East model.  A small portion of the system is
depicted horizontally at each step.  Numbers (0 and 1) indicate the
presence or absence of individual excitations in a sequence of
configurations proceeding from top to bottom.  Each of the 10 steps
comprising this schematic trajectory involves creating or (nearly)
destroying blocks of $a=8$ consecutive excitations.  (The simple
sequences of single-cell changes that create and destroy these blocks
are not shown.) This idealized example represents the lowest energy
pathway that results in elimination of the rightmost defect.}
\end{figure}

Attractions establish a length scale, $a\approx h/(h-\epsilon)$, below
which it is energetically advantageous to fill immobile spaces between
excitations with additional defects.  Gap filling (clearly visible in
the trajectory of Fig.~1) effectively renormalizes the basic
hierarchical relaxation mechanism.  Low energy pathways, such as that
depicted in Fig.~2 for $a=8$, do not involve isolated defects
separated by distances smaller than $a$.  This change of scale, which
grows monotonically with $\epsilon$,
\begin{equation}
a \approx \cases{ 1+\epsilon/h_0, & $\epsilon \ll h_0$\cr
\epsilon c^{1/2}e^{\epsilon/2}, & $\epsilon \gg h_0$,\cr}
\end{equation}
has two important consequences.  First, the activation energy for
pathways connecting subsequent excitations renormalizes.  For $\ell =
(2^m-1) a +1$, the barrier becomes $\Delta(\ell) = (m-1)h
+a(h-\epsilon) \approx h \ln{(\ell/a)}/\ln{2}$.  Second, a new time
scale $\tau_a$ emerges, corresponding to the frequency with which
excitation domains spontaneously grow to size $a$.

Because in the East model domains are bound on one side, fluctuations
of the boundaries dividing excited domains from immobile domains
(i.e., domain walls) are equivalent to the motion of a one-dimensional
random walker.  The potential energy of this walker at position $x$ is
$(h-\epsilon)x$, and a reflecting boundary restricts motion to $x>0$.
Excursions for which the change in energy is comparable to $T$ are
essentially diffusive in nature, occurring with a frequency
proportional to $(h-\epsilon)^2$.  Excursions of length
$L>(h-\epsilon)^{-1}$ are less frequent by a factor of
$\exp[(h-\epsilon)L-1]$.  The basic time scale associated with steps
of length $a\gg 1$ is therefore $\tau_a \approx (h-\epsilon)^{-2}
e^{h}$.

Assembling these results, we estimate the relaxation rate for
$\epsilon>0$ as
\begin{equation}
\Gamma \approx \left(h-\epsilon \right)^2
\exp\left[-{h} - {h\over \ln{2}}\ln(d/a) \right].
\label{equ:rate}
\end{equation}
This expression depends on attraction strength $\epsilon$ implicitly
through $h$, $d$, and $a$.  Each of these quantities grows with
increasing $\epsilon$, but their combination in Eq.~\ref{equ:rate} may
be nonmonotonic.

Through renormalization of the basic dynamical length scale,
attractions aid the propagation of excitations.  But for fixed $c$,
attractions also increase the distance excitations must propagate to
achieve relaxation.  The quantity $h \ln(d/a)$ characterizes this
competition.  For weak attractions, $\ln(d/a)\approx \ln{d_0} -
\epsilon/h_0$ varies much more rapidly than $h\approx h_0 +
2c\epsilon$, and relaxation becomes more facile with
increasing $\epsilon$:
\begin{equation}
\ln\left({\Gamma \over \Gamma_0}\right) \approx {\epsilon \over
\ln{2}}, \qquad \epsilon \ll h_0.
\label{equ:weak}
\end{equation}
Defect concentration $c$ does not appear in this result, which should
be accurate in the limit of small $c$.  Terms of order
$(\ln{c})^{-1}$, which for moderately small values of $c$ cause the
initial growth in $\ln{\Gamma}$ to differ for different attraction
strengths, have been omitted.

For strong attractions, $\ln(d/a)\approx \ln(\epsilon/c)$ varies
instead more slowly than $h\approx \epsilon$, so that $\Gamma$ decreases
with increasing $\epsilon$.  In this regime the change of
dynamic length scale is insufficient to offset the enhanced sparseness
of excitation domains.  Relaxation rates should thus be maximum
at the crossover from small-$\epsilon$ to large-$\epsilon$
behavior, i.e., near $\epsilon = h_0$.

The hierarchical mechanism we have described is only sensible for
$a<d$.  When the renormalized length scale is comparable to or larger
than $d$, there are few consecutive defects separated by distances
larger than $a$.  For $\epsilon \gtsim c^{-1}$ the minimum energy
pathway connecting defects simply excites all intervening lattice
cells.  Domain wall fluctuations are then the only relevant dynamics.
A domain wall excursion of length $d$ requires activation energy
$d(h-\epsilon)$, which approaches $c^{-1}$ in the limit of large
$\epsilon$.  Relaxation rates for $a\gtsim d$ vary with attraction
strength (at fixed $c$) in a simple Arrhenius fashion, $\Gamma \approx
[(h-\epsilon)]^{2} e^{1/c} \approx c^{-1} e^{-\epsilon+1/c}$.

Numerical simulations of the East model defined by
Eqs.~\ref{equ:hamiltonian} and \ref{equ:field} verify these expected
features.  Relaxation time $\tau$ is plotted in Fig.~3 as a function
of $\epsilon$ for several values of $c$ between 0.01 and 0.2.  We
define $\tau$ through the persistence function $\pi(t)$, i.e., the
average fraction of cells that do not change over a time interval of
length $t$\cite{berthier}.  Specifically, $\pi(\tau)=1/e$, so a
majority of cells have undergone fluctuations within time $\tau$.
Most importantly, numerical results confirm the existence of
re-entrant dynamics as attraction strength increases.  The maximum
relaxation rate can be nearly two orders of magnitude larger than its
value at $\epsilon=0$ in this range of defect concentration.  As
expected, this maximum occurs near $\epsilon = \ln(1/c)$.  Re-entrance
should become still more dramatic for smaller values of $c$.

\begin{figure}
\centerline{\includegraphics[width=8cm]{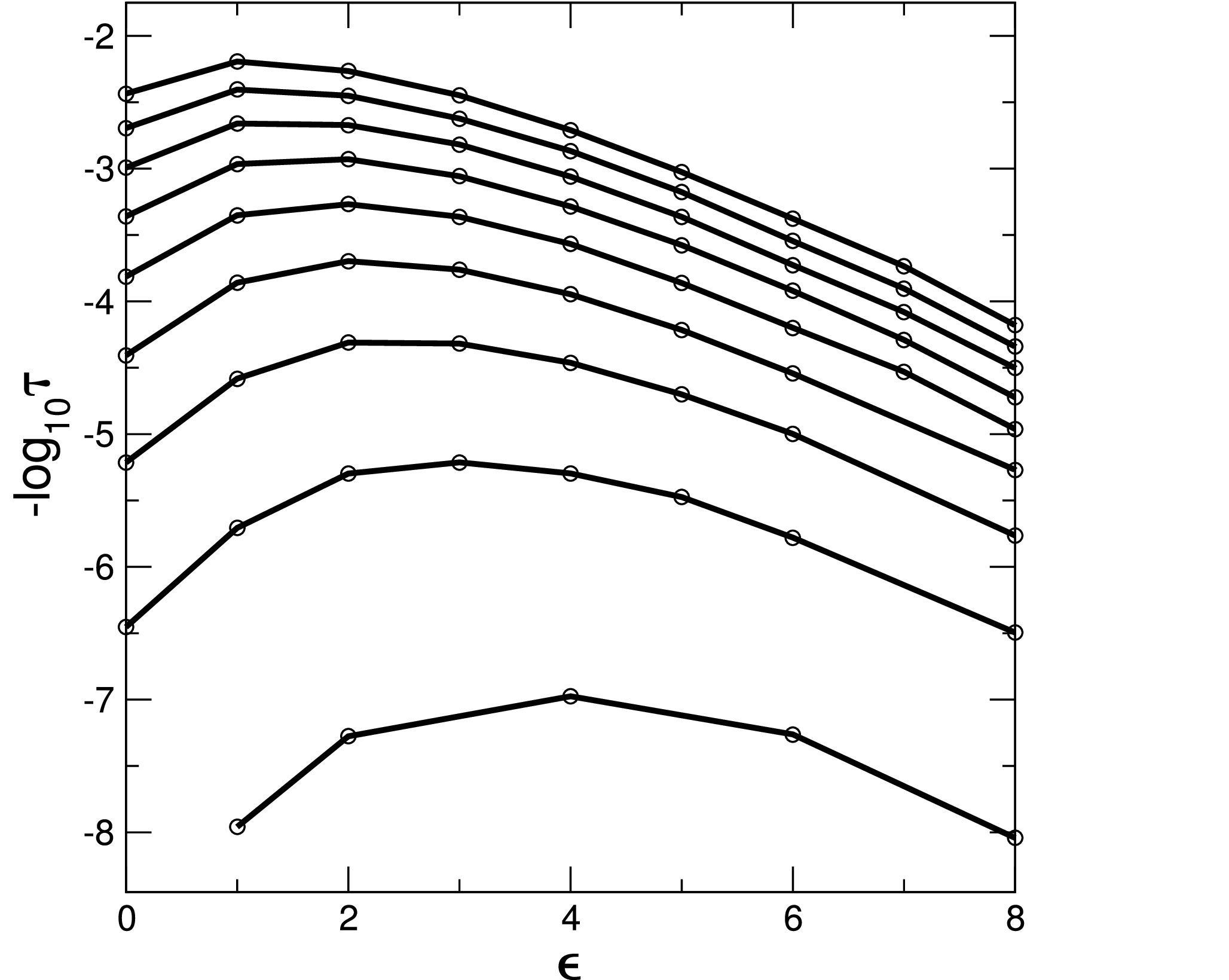}}
\caption{Average time $\tau$ required for all but $N/e$ cells to flip, as
a function of attraction strength $\epsilon$.  The concentrations are,
from bottom to top, $c=0.02, 0.04, 0.06, 0.08, 0.1, 0.12, 0.14, 0.16$,
and $0.18$. The continuous time Monte Carlo simulations that generated
these results employed $N=10^5$ lattice cells.}
\end{figure}

For the larger values of $c$ we have considered, the asymptotic
behavior for $a\gtsim d$ begins to appear in our simulation results at
large $\epsilon$.  (For smaller $c$, the corresponding range of
$\epsilon$ is numerically inaccessible.)  Several results point to
dominance of activated domain wall fluctuations in this regime.
First, the Arrhenius behavior of $\ln \Gamma$ evident in
Fig.~3 is consistent with our analysis.  Interestingly,
although the hard sphere liquid (and the $\epsilon=0$ East model) is a
fragile glass-former\cite{fragility}, our results suggest that the
re-entrant glass state is instead strong.  This prediction seems
reasonable, given that gel-like structures have much in common with
conventionally strong, network-forming materials\cite{strong}.  It
would be straightforward to test in experiments or molecular dynamics
simulations by varying temperature at fixed packing fraction and
attraction strength.

The decay of the persistence function exhibits a more direct signature
of domain wall fluctuations.  For small values of $c$, a great
majority of lattice cells are unexcited at any time.  The average
fraction of relaxed spins, $1-\pi(t)$, should thus be largely
determined by the growth of excited domains.  (Relaxation of the
initially excited minority contributes only weakly and at early
times.) Recall that the time required for a typical domain wall
excursion of length $L$ is $t \approx (h-\epsilon)^{-2}
e^{(h-\epsilon)L}$.  In other words the typical extent of domain
growth at time $t$ is $L\sim (h-\epsilon)^{-1}
\ln[(h-\epsilon)^2 t]$. As a result, the persistence function,
\begin{eqnarray}
\pi(t) &\approx& 1 - L/d
\\
&\approx& {\rm constant} - c\ln(t)
\label{equ:persist}
\end{eqnarray}
decays logarithmically. Eq.~\ref{equ:persist} suggests that the
coefficient of logarithmic relaxation for $\pi(t)$ and related
correlation functions should not change significantly with $u/T$ for
fixed packing fraction.  $\pi(t)$ is plotted in Fig.~4 for $c=0.14$
and several values of $\epsilon$.  A period of logarithmic decay, with
a slope on the order of $c$, appears as attractions are introduced and
becomes more prominent as $\epsilon$ increases.  Since the average
spacing between excited domains still exceeds $a$ in this case, the
terminal decay of $\pi(t)$ maintains a form characteristic of
hierarchical relaxation.  Logarithmic decay should extend over longer
time spans for smaller $c$ and larger $\epsilon$.  The two decades of
such decay apparent in Fig.~4 for $\epsilon \geq 6$ reflect only the
limit of our computational resources.

\begin{figure}
\centerline{\includegraphics[width=8cm]{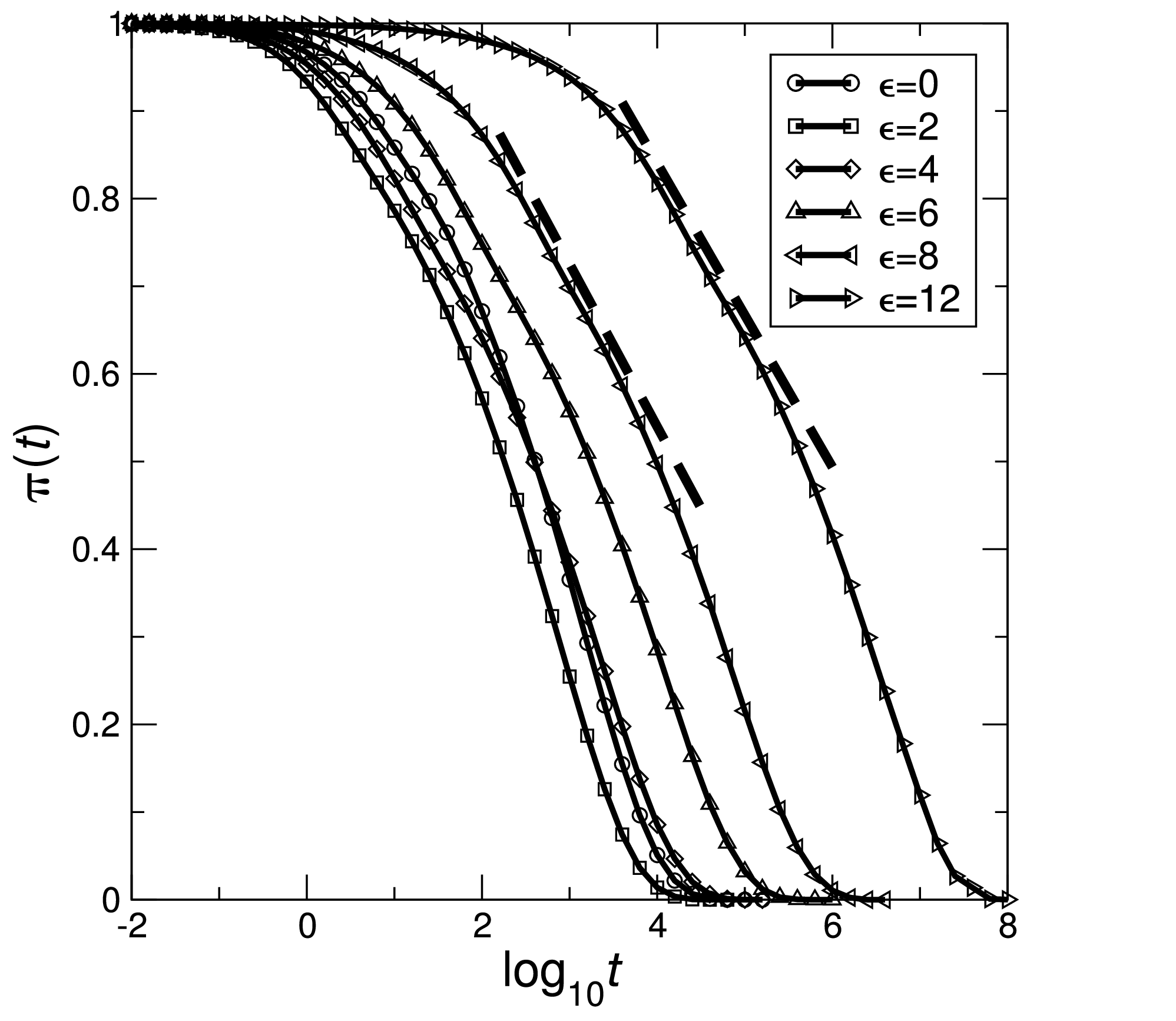}}
\caption{Persistence $\pi(t)$ as a function of time $t$ for several
values of $\epsilon$ with $c=0.14$. Dashed lines highlight regions
where logarithmic relaxation is noticeable.}
\end{figure}

In summary, the simplest physically motivated modification of the East
model reproduces the principal dynamical anomalies of attractive
colloidal suspensions.  In contrast to mode coupling theory, our model
suggests simple physical origins of these features.  It is the
combination of static and dynamic heterogeneity in our picture that
generates new dynamical features.  Clustering of mobility renormalizes
the relaxation mechanism that predominates in the absence of
attraction.  Re-entrance results directly from a competition between a
growing dynamic length scale $a$ and a growing static length scale
$d$. When the scale of uninterrupted defect propagation exceeds the
typical separation between defect domains, the primary mode of
relaxation crosses over to simple domain wall fluctuations, whose
dynamics give rise to logarithmic relaxation.  Although we have
presented results only for a modified East model, an equivalently
modified FA model shares much of the same behavior.  Basic relaxation
pathways are similarly renormalized, leading to re-entrance of
comparable magnitude.  Logarithmic relaxation is less pronounced in
this case, since excitation domains are not pinned on one side.  At
long times, fluctuations of a domain's two walls become correlated,
and our arguments for the decay of $\pi(t)$ do not strictly apply.

Based on the dynamics of the attractive East model we make several
predictions that may be tested by experiment or by numerical
simulation of colloid dynamics.  Most significantly, our results imply
for increasing attraction strength not only a transformation from
strong to fragile glass but also rapid growth of the length scale
associated with dynamic heterogeneity.  Recent computer
simulations\cite{rabani,pan} are consistent with this latter
prediction.

We are aware of another effort to rationalize the behavior of
attractive colloids within the kinetic facilitation
picture\cite{juanpe}.  This approach, in contrast to our own,
preserves a lack of static spatial correlations between defects.  It
achieves re-entrance by instead modulating the details of facilitation
constraints.  The two pictures are most distinct for very strong
attractions, where spatial correlations between defects in our model
become long-ranged.  In our view this behavior naturally reflects the
growing range of correlated density fluctuations in a disordered
material as the strength of interparticle attractions increases (or
temperature decreases).  Detailed experiments or computer simulations
should be sufficient to assess the relative validity of these
approaches by characterizing spatial distributions of dynamic
heterogeneities on short time scales.

Although the kinetic facilitation picture is most sensible at high
density, it is interesting to consider the implications of our results
for low packing fraction and very large attraction strength.  In a
version of the attractive East model in more than one dimension,
nearly irreversible aggregation of immobile regions would describe the
growth of sparse fractal patterns similar to those observed in
experiments\cite{witten,weitz2}.  It may therefore be possible to map
the entire phase diagram of our model (in the plane of $c^{-1}$ and
$\epsilon$) onto that of a real attractive colloidal suspension
(in the plane of $\phi$ and $u/T$).  

We would like to acknowledge D. Chandler, D. Fisher, and E. Rabani for
useful discussions.  We would especially like to thank J.-P. Garrahan
for useful discussions that guided some of the analytical arguments
used in this work, and D.Weitz for introducing us to this problem.  We
acknowledge the NSF (D.R.R.) and DOE (P.L.G.) for financial support.
D.R.R. is a Camille Dreyfus Teacher-Scholar and an Alfred P. Sloan
Foundation Fellow.

\bibliographystyle{prsty}

\begin{thebibliography}{10}
\bibitem{fragility} Although it is difficult to classify an athermal
material as either strong or fragile, we view hard sphere interactions
as a limit of sharp repulsion.  Dense fluids of
sharply repulsive particles, perhaps the best description of
non-attractive colloidal suspensions, do exhibit the super-Arrhenius
temperature dependence defining fragile glass-formers.  See Ref.[2].
\end{thebibliography}

\end{document}